\documentclass[global,twocolumn]{svjour}
\usepackage{psfig}
\journalname{Applied Physics A}
\begin{document}
\bibliographystyle{unsrt}
\psfigurepath{.:figure}
\title{Effect of pressure on magnetic structure in heavy fermion
CeRhIn$_5$}
\author{W. Bao\inst{1} \and S.F. Trevino\inst{2,3} \and J.W. Lynn\inst{2}
\and P.G. Pagliuso\inst{1} \and
J.L. Sarrao\inst{1} \and J.D. Thompson\inst{1} \and Z. Fisk\inst{1,4}
}                     
\offprints{W. Bao}          
\institute{Los Alamos National Laboratory, Los Alamos, NM 87544, USA 
\and Center for Neutron Research,
National Institute of Standards and Technology, Gaithersburg, MD 20899, USA
\and 
United States Army Research Laboratory, Adelphi, MD 20783, USA 
\and
Florida State University, Tallahassee, FL 32306, USA
}
\date{Received: date / Revised version: date}
%
\maketitle
\begin{abstract}
The effect of hydrostatic pressure on the incommensurate
antiferromagnetic structure 
of CeRhIn$_5$ is investigated
with neutron diffraction using a He pressure cell. 
At 3.8 kbar, the staggered magnetic moment is 0.37(4) $\mu_B$ per Ce 
at 1.6 K, which is the same as the ambient-pressure value. The N\'{e}el
temperature $T_N=3.8(1)$ K is also the same as the ambient-pressure one,
although the curve of order parameter has changed by pressure.
The incommensurability $\delta$ of the magnetic wave vector 
{\bf q}$_M=(1/2$,1/2,$\delta$) has reduced from 
$\delta=0.297$ at ambient pressure to $\delta=0.294(1)$
at 3.8 kbar.
\end{abstract}
\section{Introduction}
\label{intro}

Superconductivity and antiferromagnetism
exist in close proximity in the heavy fermion materials with
chemical formula Ce$M$In$_5$, which have Sommerfeld constants
$\gamma$= 0.4, 0.7, and 0.3 J/mole K$^2$ for
$M$=Rh, Ir, and Co, respectively\cite{hegger,joeIr}. 
These tetragonal materials (HoCoGa$_5$ structure with space group
No.\ 123, P4/mmm) consist of alternating layers of the cubic heavy fermion
antiferromagnet CeIn$_3$ and intervening $M$In$_2$\cite{nstru}.
At ambient pressure, CeRhIn$_5$ is an antiferromagnet below $T_N=3.8$ K,
with magnetic moments on the Ce ions, 0.374(5)$\mu_B$ at 1.4 K, lying in the basal plane and forming an 
incommensurate transverse spiral with a
magnetic wave vector ${\bf q}_M=(1/2,1/2,0.297)$\cite{bao00a,curro}.
Under a pressure of 17 kbar, CeRhIn$_5$ becomes a superconductor
below $T_C$=2.1 K\cite{hegger}. Both CeIrIn$_5$ and CeCoIn$_5$ are
superconductors at ambient pressure with $T_C$=0.4 and 2.3 K, 
respectively\cite{joeIr}.
Lines of nodes in the superconducting gap have been indicated from
thermodynamic, transport, and NQR\\
measurements\cite{roman,zheng,mito,izawa}.
This type of anisotropic superconductivity in heavy fermion materials
is widely believed to be mediated by
antiferromagnetic fluctuations. While a two-dimensional (2D) 
Fermi surface of
undulating cylinders is detected in de Haas-van Alphen 
measurements\cite{haga}, anisotropic 3D
antiferromagnetic correlations are observed in direct measurements using 
neutron scattering\cite{bao01b} and inferred from a theoretical fit 
to NQR measurements\cite{zheng,mito}.

Anisotropic magnetic correlation lengths of CeRhIn$_5$
indicate that the antiferromagnetic nearest-neighbor
interaction in the CeIn$_3$ layer is stronger than the magnetic interaction
between Ce neighbors that are separated by the RhIn$_2$ layer\cite{bao01b}. 
This may play a role in stabilizing the commensurate antiferromagnetic 
structure of Ce$_2$RhIn$_8$\cite{bao01a}, 
which can be viewed as a periodic stacking of 2 layers of CeIn$_3$ on a
layer of RhIn$_2$.
The incommensurate magnetic structure of CeRhIn$_5$ is robust.
While $T_N$ is reduced linearly to zero with La doping on the Ce site 
at a critical dopant concentration around 0.4\cite{pgpCeLa}, 
the magnetic structure of Ce$_{0.9}$La$_{0.1}$RhIn$_5$\\
($T_N$=2.7 K) is still characterized
by ${\bf q}_M=(1/2,1/2,0.297)$ and a staggered moment of 0.36(2)$\mu_B$
at 1.4 K\cite{bao01c}.
Applying pressure to CeRhIn$_5$ or doping it
with Ir on the Rh site has only a small effect
on $T_N$ until the material becomes a superconductor\cite{hegger,mito,pgpRhIr}.
We have found with neutron diffraction measurements that the 
suppression of the
antiferromagnetic phase by Ir doping is through progressive
reduction of the staggered moment of the incommensurate magnetic 
spiral\cite{bao01e}. Here we report the 
effect of pressure on magnetic structure of CeRhIn$_5$.

\section{Experiments and Results}
\label{sec:1} 

High pressure neutron diffraction experiments were performed at NIST
using the thermal triple-axis spectrometer BT2 in a two-axis mode.
To reduce neutron absorption by In and Rh, neutrons of
incident energy $E=35$ meV were selected using the (002) reflection of a pyrolytic graphite (PG) monochromator. A PG filter of 5 cm thickness was inserted in the incident neutron beam to remove higher order neutrons.
The horizontal collimations were 60-40-40.

The pressure cell\cite{bao94b} was made of a BeCu alloy, Berylco 25. 
The cell body is a cylinder with outer diameter 1/2 inch and 
inner diameter 1/8 inch. Helium was used as the pressure transmitting medium
and was compressed into the pressure cell through a stainless steel capillary.
The pressure in the cell was monitored by a manganin resistance gauge and
by measuring the lattice constant of graphite in the cell.
During the experiment at 3.8 kbar,
pressure decreased 0.04 kbar due to small He leaks. 

The single crystal sample of CeRhIn$_5$ was grown from an In flux. 
It was cut to a 
rectangular bar to fit inside the pressure cell so that the axis of
the cell was parallel to the (1,$-1$,0) crystal orientation. The neutron
scattering plane was the ($hhl$) plane. The pressure cell was mounted
on the cold finger of a top loading, pumped He cryostat.

Solid circles in Fig. 1 show a pair of magnetic Bragg peaks in 
\begin{figure}
\centerline{
\psfig{file=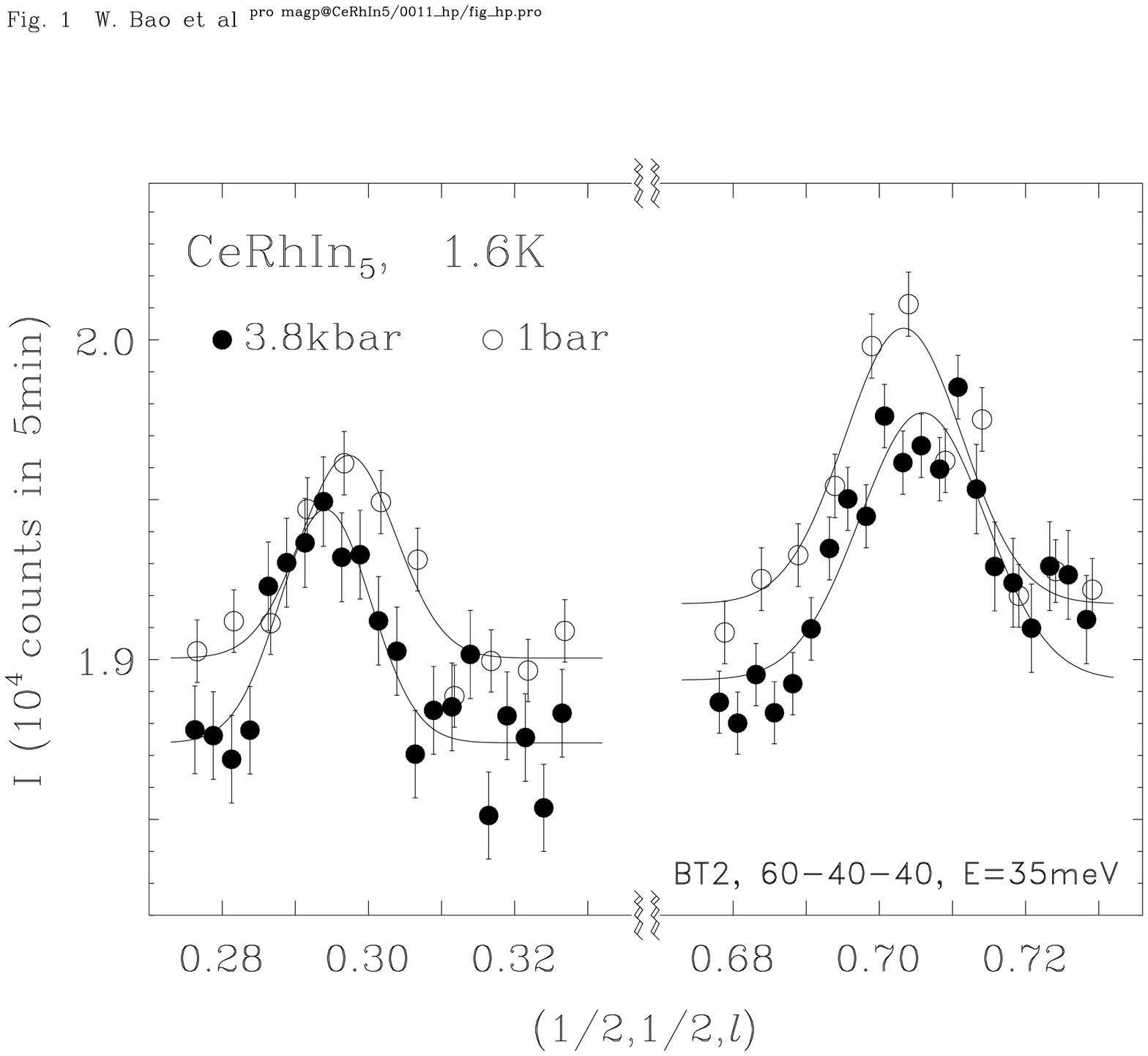,width=\columnwidth,angle=90,clip=}}
\caption{Elastic scan through a pair of magnetic Bragg points 
at 1.6 K and under pressure of 3.8 kbar (solid circles) and 1 bar (open
circles).  
} 
\end{figure}
\label{fig:1}       
a Brillouin zone, measured at 3.8 kbar. No other peaks were found
along the (1/2,1/2,$l$) line in a search from (1/2,1/2,0) to
(1/2,1/2,1). Compared to data
measured at 1 bar (see open circles), it is clear that the
period of the incommensurate spiral increases with pressure.
The magnetic wave vector, (1/2,1/2,$\delta$), changes from
$\delta=0.297(1)$\cite{bao00a} at ambient pressure to 0.294(1) at 3.8 kbar.
Intensities of the magnetic Bragg peaks at the two pressures,
however, remain the same within the
error bars. The staggered moment is determined to be 0.37(4) $\mu_B$
per Ce ion by comparing magnetic Bragg intensities of the 3.8 kbar and
1 bar measurements.

The intensity of the (1/2,1/2,0.706) magnetic Bragg peak is shown in
Fig. 2 as the square of the order parameter of the magnetic phase transition
\begin{figure}
\centerline{
\psfig{file=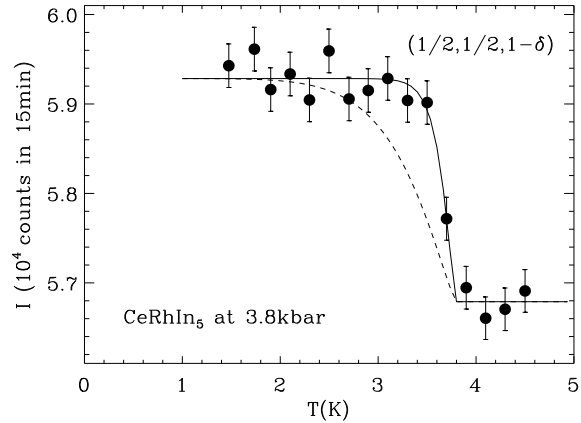,width=\columnwidth,angle=90,clip=}}
\caption{Temperature dependence of the (1/2,1/2,0.706) magnetic
Bragg peak of CeRhIn$_5$ under 3.8 kbar pressure. 
The solid line is a guide to the eyes, and the dashed line represents
the square of the order-parameter curve at 1 bar\protect\cite{bao00a}.
} 
\end{figure}
\label{fig:2}       
at 3.8 kbar. The N\'{e}el temperature changes little, which is consistent
with bulk and NQR measurements\cite{hegger,mito}. 
However, the intensity increases more rapidly below
$T_N$ under pressure as compared to the ambient pressure result.

Mito and co-workers recently reported a linear reduction with pressure 
of the internal field at the In(1) site in an NQR study on CeRhIn$_5$\cite{mito}.
They suggested that either the staggered moment decreased with pressure or
the moment tilted progressively towards the $c$-axis when pressure was
raised to the critical pressure of 16.3 kbar. The former explanation would
imply a $\sim$25\% reduction of the
staggered moment at 3.8 kbar and is not consistent with our data.
The latter explanation may lead to extra magnetic Bragg peaks characterized
by a wave vector (1/2,1/2,0) or (1/2,1/2,1/2), which we do not observed in
our work. One possibility is that the hyperfine interaction at the In(1) site
may be sensitive to pressure.

We would like to thank N.J. Curro and T. Mito for useful discussions.
Work at Los Alamos was performed under the auspices of the U.S. Department
of Energy. P.G.P. also acknowledges FAPESP-SP (Brazil).

%


\begin{thebibliography}{00}

\bibitem{hegger}
{ H. Hegger, C. Petrovic, E.G. Moshopoulou, et al., Phys. Rev. Lett. {\bf 84}, 4986 (2000)}.

\bibitem{joeIr}
{ C. Petrovic, R. Movshovich, M. Jaime, et al., Europhys. Lett. {\bf 53}, 354 (2001);
 C. Petrovic, P. G. Pagliuso, M.F. Hundley, et al., J. Phys. Condens. Mat. {\bf 13}, L337
  (2001)}.

\bibitem{nstru}
{ E.G. Moshopoulou, Z. Fisk, J. L. Sarrao and J. D. Thompson,
J. Solid State Chem. {\bf 158}, 25 (2001)}.

\bibitem{bao00a}
{ W. Bao, P. G. Pagliuso, J. L. Sarrao, et al., 
  Phys. Rev. B {\bf 62}, R14621 (2000); Erratum: ibid. {\bf 63}, 219901(E) (2001)}.

\bibitem{curro}
{ N.J. Curro, P.C. Hammel, P.G. Pagliuso, et al., Phys. Rev. B {\bf 62}, R6100 (2000)}.


\bibitem{roman}
{ R. Movshovich, M. Jaime, J.D. Thompson, et al., Phys. Rev. Lett. {\bf 86}, 5152 (2001)}.

\bibitem{zheng}
{ G.-Q. Zheng, K. Tanabe, T. Mito, et al.,
  Phys. Rev. Lett. {\bf 86}, 4664 (2001)}.

\bibitem{mito}
{ T. Mito, S. Kawasaki, G.-Q. Zheng, et al., Phys. Rev. B {\bf 63}, 220507(R) (2001)}.

\bibitem{izawa}
{ K. Izawa, et al., cond-mat/0104225 (2001)}.

\bibitem{haga}
{ Y. Haga, Y. Inada, H. Harima, et al., Phys. Rev. B {\bf
  63}, 060503(R) (2001);
 A.L. Cornelius, A.J. Arko, J.L. Sarrao, et al., Phys. Rev.
  B {\bf 62}, 14181 (2000);
 D. Hall, et al., cond-mat/0011395 (2000);
 D. Hall, et al., cond-mat/0102533 (2001)}.

\bibitem{bao01b}
{ W. Bao, G. Aeppli, J. W. Lynn, et al., cond-mat/0102503
  (2001)}.

\bibitem{bao01a}
{ W. Bao, P. G. Pagliuso, J. L. Sarrao, et al.,
   Phys. Rev. B
  {\bf 64}, 020401(R) (2001)}.

\bibitem{pgpCeLa}
{ P.G. Pagliuso, et al., unpublished (2001)}.

\bibitem{bao01c}
{ W. Bao, A.D. Christianson, P.G. Pagliuso, et al., cond-mat/0109379
   (2001)}.

\bibitem{pgpRhIr}
{ P.G. Pagliuso, C. Petrovic, R. Movshovich, et al., cond-mat/0101316 (2001)}.


\bibitem{bao01e}
{ A.D. Christianson, W. Bao, et al., unpublished (2001)}.

\bibitem{bao94b}
{ W. Bao, C. Broholm and S. F. Trevino, 
Rev. Sci. Instrum. {\bf 66}, 1260 (1995)}.

\end{thebibliography}
\end{document}